# Las representaciones del cielo entre los Tomárâho

*Guillermo Sequera*[1] *y Alejandro Gangui*[2]

## Abstract

The small community of the Tomárãho, an ethnic group culturally descendant from the Zamucos, became known on the South American and international scene only recently. Hidden from organized modern societies until the late 1980s, the Tomárãho maintained close contact with nature and developed original ways of explaining and understanding it. This article presents the first results of an interdisciplinary project seeking to provide a detailed analysis of different astronomical elements in the imagined sky of the Tomárãho. After a brief history of the community and their present day lifestyle, we introduce some of the protagonist figures of their ethnoastronomy. The main body of the article focuses on the stratification of the sky after the collapse of the chãro tree which served as the world's axis in the Tomárãho cosmology. The study reveals the richness of the Tomárãho culture and highlights the need for more detailed research into the names and stories of the visible astronomical bodies and phenomena which have been observed by the inhabitants of this part of the Paraguayan Chaco.

---

1   Secretaría Nacional de Cultura de Paraguay
2   Instituto de Astronomía y Física del Espacio, Conicet/Universidad de Buenos Aires, Argentina

## Resumen

La pequeña comunidad de los Tomárãho, un grupo étnico cultural descendiente de los Zamucos, llegó a ser conocido en la América del Sur y en la escena internacional sólo recientemente. Oculto en las sociedades modernas organizadas hasta finales de 1980, los Tomárãho han mantenido un estrecho contacto con la naturaleza y desarrollado formas originales de la explicación y la comprensión de la misma. Este artículo presenta los primeros resultados de un proyecto interdisciplinario para proporcionar un detallado análisis de los diferentes elementos astronómicos en el imaginario cielo de los Tomárãho. Después de una breve historia de la comunidad y su estilo de vida actual se presentan algunas de las figuras protagonistas de su etnoastronomía. El cuerpo principal del artículo se centra en la estratificación del cielo, después de la caída de la châro, árbol que sirvió como eje del mundo en la cosmología Tomárãho. El estudio pone de manifiesto la riqueza de la cultura Tomárãho y pone de relieve la necesidad de una investigación más detallada sobre los nombres y las historias de los cuerpos visibles y fenómenos astronómicos que han sido observados por los habitantes de esta parte del Chaco paraguayo.

## Palabra clave

Etnoastronomía, antropología, chamacoco, Alto Paraguay

La pequeña comunidad de los Tomárâho, etnia del tronco cultural de los Zamucos, irrumpió en el escenario de la antropología sudamericana y mundial hace relativamente poco tiempo. Siempre alejados de las riberas del río Paraguay, este pueblo permaneció por muchos años oculto a la mirada de las sociedades organizadas modernas. Como todo pueblo en estrecho contacto con la naturaleza, los Tomárâho mostraron una profunda y rica cosmovisión, que guarda paralelos con otras culturas aborígenes quizás más estudiadas, como así también detalles que le son propios. Esto también se da en su imaginario del cielo y de los personajes que guardan una estrecha relación con la bóveda celeste. Este trabajo se basa en los prolongados estudios antropológicos de G.Sequera quien, recientemente y en colaboración con el astrofísico A.Gangui, ha dado inicio al proyecto de llevar a cabo un análisis pormenorizado de los diversos elementos astronómicos presentes en el cielo imaginado por los Tomárâho y por otras etnias Chamacoco. Se hará una breve reseña sobre esta cultura aborigen: lugares donde viven y regiones de influencia en el pasado, su familia lingüistica, su modo de vida y cómo el avance de "la civilización" afectó su cultura (y su supervivencia). Mencionaremos luego los trabajos de campo ya realizados durante décadas y algunos estudios y publicaciones existentes. Haremos también una descripción de la metodología de este trabajo de campo con sus prácticas antropológicas particulares (singularidades y dificultades). Por último, nos concentraremos en la forma en que conciben el cielo los Tomárâho, y describiremos el análisis que hemos venido haciendo en los últimos meses de trabajo conjunto.

## La cultura Tomárâho

El acto de vivir de una cultura se muestra de muchas maneras diferentes. Una posible es esa creatividad puesta en juego por el grupo humano en observar los fenómenos cambiantes de la naturaleza en su entorno, ponerles nombres a las cosas, nominar a las personas, a los animales y a las plantas, en fin, construir conocimientos y símbolos propios que le permitan adquirir una identidad y sobrevivir como sociedad. A esto aspiran los pueblos actuales y, en completo paralelo, fue una motivación fuerte de las culturas del pasado; de aquellas más desarrolladas tecnológicamente, como así también de los pueblos que no siguieron el paso frenético de la "civilización" y permanecieron en un contacto más estrecho con la naturaleza. La pequeña comunidad indígena de los Tomárâho es un ejemplo cercano de estos últimos.

Un primer contacto con representantes de este pueblo se dio en febrero de 1986, ocasión en la que uno de los autores (Sequera) pudo comprobar la situación de semi-exclavitud y fuerte dependencia que debían soportar los integrantes de este pueblo en su relación con la Empresa Carlos Casado. En complicidad con el Estado Paraguayo y desde finales del siglo XIX (pocos años después de terminada la guerra de la Triple Alianza que debilitó sobremanera esa nación), esa empresa de capitales anglo-argentinos se había apropiado indebidamente de las tierras de los Tomárâho y explotaba los quebrachales en forma intensiva con el fin de extraer madera para durmientes de vías férreas y para abastecer a sus industrias tanineras. Sin ningún tipo de derecho reconocido, los nativos se veían obligados a trabajar como hacheros bajo duras condiciones laborales y a ser testigos de la expoliación de los bosques chaqueños, convirtiéndose en víctimas indefensas del avasallamiento de su espacio vital (Sequera, 2006).

Reconocida la importancia de llevar adelante un estudio antropológico en el caso de esta minoría étnica, se realizaron estudios comparativos de fuentes etnográficas en diversas bibliotecas y centros de investigación, que ayudaron a clarificar el panorama sobre el posible origen y mención de los Tomárâho en cartas anuas jesuíticas y crónicas antiguas. Sequera, además, convivió con los nativos durante varios períodos entre 1987 y 1992, lo que le permitió llevar adelante un trabajo metódico de reconocimiento y transcripción de la lengua Tomárâho, el inventariado en detalle de la representación social de su flora y su fauna, así como también la colecta y registro de un inmenso corpus mítico que trasluce la rica cosmovisión de esa parcialidad étnica. El trabajo etnográfico llevado adelante empleó las más adecuadas técnicas cualitativas de la antropología, con rigor científico en las observaciones y otorgándole suma importancia al trabajo de campo, basado en la actitud metodológica de la observación participante, de acuerdo con los trabajos del antropólogo polaco Bronislaw Malinowski (Malinowski, 1922).

Los Tomárâho (o Tomaráxo) son un pequeño grupo étnico que, junto con los Ybytóso (o Ebidóso), forman el grupo más grande de los Ishir (este último grupo es conocido en el Paraguay como los Chamacoco). Tradicionalmente cazadores-recolectores, los Chamacoco son ubicados en la clasificación lingüística como emparentados a la familia Zamuco. Otros grupos indígenas, los Ayoreo, también pertenecen a la familia lingüística Zamuco [ver Fig.1]. No es nuestra intención entrar aquí en las discusiones sobre la etimología de estas palabras y denominaciones, donde abundan propuestas dispares basadas principalmente en los escritos de cronistas europeos, comenzando quizás con el Viaje al Río de la Plata, 1534-1554 del arcabucero Ulrico Schmidl (1510-1580). En diversos catálogos de lenguas y dialectos, en recopilaciones de crónicas y otros textos europeos, surgen términos como Timinabas, luego Timinaha y otros similares, siempre en la categoría de los Zamuco. Se menciona que los pueblos que hablaban estas lenguas se hallaban en los bosques del Chaco, lejos del río Paraguay, hacia el interior de la región. Se menciona, además, que estos indígenas aun no habían sido "reducidos a la misión jesuítica". Todo da a pensar, pues, que se están refiriendo a los Tomárâho, aquellos habitantes cuyos descendientes Sequera visitó en proximidades de la zona

de San Carlos en 1986 y que, por sus costumbres ancestrales e idiosincrasia propias, permanecieron por muchos años al margen de la sociedad paraguaya [Fig.2].

## Los primeros contactos

Los Tomárâho fueron reconocidos y observados por diversos estudiosos ya a partir de fines del siglo XIX. Guido Boggiani, artista plástico italiano que se instala en Paraguay para la explotación del quebracho colorado, dedicó su tiempo libre en aprender sobre las costumbres de los Chamacoco. Instalado en Puerto Pacheco en el alto Chaco, Boggiani distingue dos categorías de indígenas, los "mansos" (que hoy conocemos como los Ybytóso) y los "bravos" (los Tomárâho). Estudió las costumbres y la lengua de los primeros, pero no alcanzó a tomar contacto con la comunidad de los últimos. El sueño de llegar a ellos y colectar de primera mano los datos etnográficos que aspiraba se vio truncado con la muerte del explorador, en plena misión, y envuelta en un halo de misterio que duraría hasta el día de hoy. A Boggiani siguió el alemán Herbert Baldus en 1923, quien retomó y extendió los estudios del italiano sobre las divisiones de los Chamacoco en tribus y sus ubicaciones geográficas y demás datos demográficos (Baldus, 1932). En sus escritos el autor señala que, en ese entonces, la población de los Tomárâho estaba compuesta por unas 300 familias.

Luego de los trabajos de Baldus habrá que esperar tres décadas para que los Chamacoco vuelvan a ser el foco de interés de antropólogos y lingüistas, esta vez en los trabajos de la investigadora eslovena Branislava Susnik. Es a partir de 1957 que Susnik inicia sus trabajos de campo y de colecta de datos entre los Ybytóso que viven y trabajan en empresas forestales ubicadas a lo largo del río Paraguay. Comparando con estudios previos y de mano de sus propias observaciones, Susnik percibe el avasallador impacto de los nuevos modos de vida de la región sobre la cultura Ybytóso. La modificación de su habitat modifica sus costumbres, sus condiciones de vida, su régimen alimenticio y su antigua división del trabajo comunitario (de la "caza al trabajo"). También surge un nuevo criterio de posesión y propiedad en el pasaje de la "búsqueda a la apropiación de bienes y cosas", y se modifican asimismo otros valores tradicionales, por ejemplo el de la caza, cuando la introducción de la escopeta desplaza la ancestral relevancia que tenía la destreza en el empleo de las armas indígenas. Por otra parte surgen conflictos entre jóvenes y ancianos que ponen en peligro la antigua transmisión de la sabiduría Chamacoco entre generaciones, lo que erosiona fuertemente los cimientos de la etnia y da pie a una transculturación acelerada (Susnik, 1995 [1969]).

## Las prácticas chamánicas

El chamanismo entre los Chamacoco tiene una gran relevancia, en todo similar a la que encontramos en otras culturas indígenas. La música vocal está muy relacionada con los rituales de los chamanes, ya se trate de hombres o de mujeres. Estos integrantes prominentes de la tribu, conocidos como konsaha o ahanak, construyen sus propios repertorios basándose en sus sueños, llamados chykêra, los que estimulan la creación de los textos poéticos, las melodías y los ritmos. Los chamanes Chamacoco buscan dominar los sueños para proyectarlos en un canto, al que se denomina teichu. La producción de estos cantos es eminentemente personal, y puede ser transmitida a otros miembros de la comunidad, pero su ejecución es en general grupal, donde se mezclan varios cantos y se crea una atmósfera sonora muy atípica, con acompañamiento de instrumentos como sonajas y silbatos.

Las sonajas, llamadas por los Chamacoco osecha o paîkâra, son construidas a partir de las calabazas de algunas plantas trepadoras (de la especie Lagenaria siceraria) o con caparazones de tortugas terrestres, llamadas enermitak (Geochelone carbonaria), donde se introducen semillas secas o piedritas que les dan sonoridad. Estas sonajas vienen a representar el cielo y los chamanes identifican su parte superior con el centro de la bóveda celeste, porn hosýpyte [Fig.3]. El cuerpo del instrumento está pintado con "relatos visuales" en forma de rombos y en el interior de estas figuras geométricas se representan las estrellas porrebija (Sequera, 2006). Los estudios sugieren una estrecha relación entre las técnicas vocales e instrumentales de los Chamacoco -especialmente de los Tomárâho, donde el chamanismo ha perdurado más intensamente- y su visión del mundo, un lazo entre la expresión sonora y la cosmovisión indígena que aun queda por investigar en profundidad (Cordeu, 1994).

Las prácticas chamánicas de los Chamacoco guardan muchas similitudes con las de otras culturas indígenas sudamericanas, e incluso de otros continentes: los sueños visionarios, el desdoblamiento de la personalidad, el trance alcanzado a través del canto, etc, son situaciones que se repiten en espacio y en tiempo. Es a través de estos sueños que los konsaha descubren para el resto de los integrantes de la tribu la verdadera "topografía" del universo indígena y su interrelación con los relatos míticos heredados de sus ancestros. (Una exploración antropológica muy detallada, pero para el caso de los Qom, o tobas del Chaco argentino, puede verse en (Wright, 2005).)

## La imagen del mundo

Aunque los detalles topográficos del cosmos imaginado por los chamanes puede diferir entre comunidades indígenas vecinas, e incluso más entre culturas separadas por grandes distancias y períodos históricos, muchas veces su estructura fundamental es muy similar. En general, existe una región superior que interpretamos como el cielo. Los hombres habitan el mundo del medio, que casi siempre es imaginado como una superficie plana rodeada por grandes extensiones de agua. Por último, el tercer reino, un mundo sumergido y oscuro, se ubica por debajo de la tierra y en él se penetra por ciertos lugares singulares, como grutas y grietas del terreno. En general, cada uno de estos reinos está habitado por espíritus que se relacionan con los hombres de muy diversas maneras. Las almas de hombres poderosos de las tribus, por lo general la de los chamanes, son capaces de viajar por estas regiones durante los sueños o los trances a los que llegan en ocasión de las ceremonias (Krupp, 1996). Durante los trances, estos representantes y defensores de las tribus se encuentran con espíritus benévolos, con los que deben interactuar, y también deben hacerles frente a los demoníacos, que pretenden llevar el mal a toda la tribu. Por ello, son frecuentes las luchas contra estos últimos, en las que solo aquellos chamanes con grandes poderes pueden resultar victoriosos.

El radio de influencia de una dada cultura primitiva puede denominarse su microcosmos. Este es necesariamente un espacio cerrado y limitado por regiones desconocidas. Hacia el interior de las "murallas" (reales o imaginadas) se ubica ese espacio familiar, organizado y habitado, con todos los elementos que caracterizan un cosmos (Eliade, 1998 [1957]). Por el contrario, hacia el exterior de esos bordes se extienden las regiones oscuras, relacionadas con "los otros", quizás con los extranjeros temibles, incluso con los peligros y la muerte. En todo espacio vital limitado existe algún lugar singular, especial por las características que la población le ha conferido. Uno de los sitios particulares que más relevancia ha tenido para muchas culturas es lo que se conoce como el "centro". Culturas avanzadas y otras más primitivas han coincidido en este lugar común, que excede las lenguas y las costumbres. Así por ejemplo, entre los habitantes de la antigua ciudad-estado de Babilonia, el zigurat (como el dedicado al dios Marduk) ocupaba dicho centro mientras que, para los antiguos hebreos, hacía lo propio el Monte del Templo en la ciudad vieja de Jerusalén; por su parte, la Kaaba o "casa de Dios" en La Meca, ciudad natal de Mahoma, es el equivalente del centro del mundo para los musulmanes. Muchos otros ejemplos han sido estudiados en la literatura (Krupp, 1996).

En el imaginario de los pueblos antiguos, estos sitios no solo representaban el centro del cosmos; también allí había ocurrido la Creación y de esos lugares había surgido la vida. Eran estos los puentes naturales, los enlaces verticales privilegiados, entre el cielo y la tierra y los únicos lugares desde donde la comunicación entre ambos reinos era posible. Podemos decir que el poder de los cielos y el efecto renovador de la potencia celestial descendía por los laterales de estos ejes simbólicos.

## Elementos astronómicos del cielo Ybytóso y Tomárâho

Los Chamacoco imaginan el mundo como una superficie plana en forma de disco, al que llaman hñymich. Sobre este inmenso disco de tierra se ubican los paisajes familiares para ellos, sus pueblos, ríos y bosques, y también las comarcas vecinas de otros pueblos con los que tuvieron contacto. El hñymich se apoya sobre las aguas de un mundo acuático al que llaman niogorot urr. Como sucede con otros pueblos antiguos, la presencia de un mundo acuoso subterráneo entra dentro de la cosmovisión Chamacoco por la importancia que ellos asignan a las fuentes de agua, en pozos y ríos (además, los seres míticos ahnapsûro son seres acuáticos, como veremos). El niogorot urr está subdividido en varios estratos, ubicados a diferentes profundidades. Por encima del disco de la tierra se ubican varios cielos transparentes, llamados genéricamente porrioho. Son semiesferas inmensas que rodean a los hombres y que ellos imaginan apoyadas en los bordes del disco hñymich.

Aparte de estas descripciones relacionadas con la tierra, el inframundo y los cielos característicos, la concepción Chamacoco de lo celeste incluye muchas representaciones relacionadas con las estrellas, tanto individuales como en grandes grupos. Encontramos la Vía Láctea, llamada iomyny, y su significación, quizás como camino de las almas (Giménez Benítez et al., 2002), y otras "nebulosas" del cielo bien visibles desde las regiones del gran Chaco Paraguayo, como las Nubes Grande (kajywysta) y Pequeña (kajywyhyrtâ) de Magallanes. También el Sol, al que llaman Deich, y la Luna, Xekulku, son protagonistas de varios relatos caros a su cultura (Cordeu, 1990-1991). Existen también ciertas narraciones que involucran al astro Venus, que ellos llaman Iohdle, o también madre de las estrellas, porrebe bahlohta, y que guardan relación con los demás astros y con los gentiles.

Muchos elementos del cosmos Chamacoco tienen seres protectores asignados, ya de trate de insectos (protegidos por el ser Ñiogogo, la rana Bufo granulosus) o de otros individuos de la avifauna local (protegidos por Wohôra, para los Tomárâho, o por Pêeta yrâhata, para los Ybytóso). Por su parte, las estrellas son protegidas por Abich, el astro hijo de Venus, mientras que el ñandú o avestruz local (Rhea americana), llamado Pemme-Kamytêrehe por los Chamacoco, se encarga del cuidado de Deich, el Sol [Fig.4]. En sus relatos y dibujos se ha podido representar también el kululte o châro, que es el "árbol cósmico" o sostén del mundo y que, como en muchas otras culturas, funciona como nexo entre el porrioho y el niogorot urr. Como veremos, el châro tiene un lugar privilegiado en las narraciones de los viajes chamánicos.

## El axis mundi

El simbolismo del centro está presente en muchas culturas y fue profundamente estudiado por los historiadores de las religiones. De acuerdo a Eliade la existencia de un centro cósmico es una consecuencia natural de la división de la realidad en lo sagrado (donde se concentra todo el valor) y lo profano (cuyo espacio no da ninguna orientación al hombre) (Eliade, 1998 [1957]). El mundo adquiere así un significado solo a través de hierofanías (del Griego, hieros, que indica sagrado, y phainein, revelarse, esto es, hierofanía podría traducirse como "donde se revela lo sagrado"). Estas intrusiones de lo sagrado en lo profano establecen un sitio singular, un "centro", que rompe con lo homogéneo de un espacio no relacionado con ninguna herencia mítica. Se interpreta también como un elemento que hace de nexo entre diversos planos existenciales; entre estos, uno es por supuesto el de la vida ordinaria, mientras que los otros niveles no son accesibles para los hombres, o al menos no para todos.

La disposición espacial de estos diferentes niveles se orienta en forma ortogonal al espacio plano terrestre, si bien es cierto que los límites de la tierra habitable también tienen connotaciones especiales. El traspaso entre estos niveles se da en la dirección vertical, ya sea hacia lo alto o, por el contrario, adentrándose en las entrañas de la tierra. La imagen típica que surge en el imaginario de los pueblos de diversas culturas es la de una montaña prominente, que se destaca en el paisaje, o la de un "árbol cósmico", notable por su envergadura o ancianidad, u otro "pilar" que cumple la función de unir el cielo con la tierra y con las regiones inferiores.

A este objeto se lo denomina en forma genérica un axis mundi, y la historia es pródiga en ejemplos; el zigurat mencionado previamente entre los babilonios, o el Monte Meru, la montaña sagrada de la tradición mitológica Hindú, por solo mencionar un par de casos. El zigurat representaba una montaña cósmica y el sacerdote que ascendía por sus terrazas accedía a la cima del universo; sus siete divisiones verticales emulaban los siete cielos astronómicos. También los templos eran construidos a imagen del axis mundi y unían diferentes niveles cósmicos, como es el caso del enorme monumento budista Borobudur ubicado en la isla de Java, en Indonesia. Llegar a su terraza superior equivalía a abandonar el espacio profano y a elevarse por la región pura de un nuevo nivel existencial (Eliade, 1974 [1955]). Veremos a continuación que existe un fuerte paralelo entre estas creencias y la figuración del cosmos de los Ybytóso y los Tomárâho del alto Chaco Paraguayo.

La montaña o pilar cósmico no solo se hallaba localizada en el centro del espacio organizado de las comunidades arcaicas; también muchas veces su cima representaba el sitio más elevado del mundo, una zona que ni los mayores diluvios de la humanidad habían podido alcanzar. Estos sitios fueron imaginados, a su vez, como una suerte de ombligo de la Tierra: lugares donde todo había sido creado, como si se tratase de un embrión. La Creación del mundo allí tomaba su lugar, para luego ir expandiéndose hacia la periferia, en todas las direcciones. Y por supuesto, también el hombre había tenido su génesis en ese centro del mundo; un centro de la Creación que un futuro trabajo etnográfico deberá también precisar para la comunidad de los Chamacoco, y para los Tomárâho en particular.

## El axis mundi Ybytóso y Tomárâho

La visión del universo aceptada por los Chamacoco imagina un árbol-sostén del mundo al que llaman kululte o châro. Este árbol pertenece a la especie Chorisia insignis, endémica del Paraguay y de países limítrofes; especie conocida en Castellano, entre otras denominaciones, como palo borracho. Como en muchas otras culturas, este árbol cósmico representa el nexo entre el cielo y la tierra. Se dice que es en la base de dicho árbol mitológico donde convergen todas las sepulturas [Fig.5].

El universo superior de los Chamacoco es imaginado como una superposición de cielos transparentes, que algunos párrafos más arriba llamamos porrioho, opuesto a la morada de los muertos en la tierra, donde hunde sus raíces el châro. El mito de origen narra que, en los tiempos de los antiguos, el cielo y la tierra estaban unidos por el árbol cósmico. Ambos reinos se encontraban, por así decirlo, fusionados, y los gentiles podían recorrer todo el ancho de estos lugares sin barreras ni impedimentos. Los primeros habitantes de la tierra, llamados yxyro poruwuhle, se alimentaban sin esfuerzo alguno, capturando animales y recogiendo los frutos de la tierra con facilidad.

Esta situación recuerda épocas mitológicas similares en otras culturas y bien puede nominarse una suerte de paraíso o jardín de la abundancia de los Chamacoco. Sin embargo, como los informantes que colaboraron con la investigación etnográfica coinciden en relatar, la historia cambió su rumbo un día en el que una viuda y sus hijos pidieron asistencia para proveerse de comida, ayuda que les fue denegada. Al ver esa falta de altruismo para con su familia y la pereza que mostraban sus vecinos, la viuda, llamada Dagylta, tomó la forma de un escarabajo y lentamente comenzó a roer la madera del poderoso châro. En ese momento aparece en la historia un pájaro, el dichikîor, de la especie Polyborus plancus o Caracara plancus, conocido en Castellano como carancho, que intentó impedir a Dagylta llevar a cabo su cometido. Pero no lo logró, y finalmente el árbol cósmico terminó desplomándose sobre la tierra. (Entre los mocovíes de la zona chaqueña argentina existe una historia muy similar, donde la mujer se habría transformado en carpincho. Véase la crónica del padre jesuita Guevara, 1969 [1764]).

Dos imágenes que presentamos aquí representan la concepción del árbol cósmico en dos registros gráficos del integrante de la comunidad ybytóso Ogwa Flores Balbuena (Sequera, 2005). La primera (Fig.5) muestra una representación del mito de origen de la comunidad ybytóso y el lugar central que en éste ocupa el châro. En el dibujo puede distinguirse un extenso número de representantes de la avifauna local, así como también la acción de los gentiles que se desplazan por el tronco principal del árbol que conecta el cielo y la tierra. El segundo registro gráfico (Fig.6) muestra el árbol Ebyta (otro nombre del châro) como el pilar y sostén del mundo, en épocas previas a su caída por la accion de Dagylta. Este dibujo muestra los dos reinos unidos por el poderoso árbol y nuevamente algunos personajes que transitan por él.

Mientras Dagylta, bajo la forma de un escarabajo, desgastaba y debilitaba el tronco del árbol cósmico, muchos gentiles, que hasta entonces circulaban libremente entre cielo y tierra, previendo lo que podría suceder, decidieron descender. Otros, en cambio, más perezosos, quedaron rezagados y, una vez vencido el árbol, permanecieron por siempre en el reino de arriba, aferrados al firmamento, y se convirtieron en porrebija, las estrellas que pueblan el cielo de los Chamacoco. (Notemos que los Bosquimanos del desierto de Kalahari, en el sur del África, cuentan que las estrellas son los primeros pueblos del mundo y que son como ellos, nómades, cazadores y recolectores; ver (Krupp, 1996). Por su parte, para los Qom del Chaco argentino, luego de un cataclismo que alteró la estructura del cielo y de la tierra, ciertos personajes también se convirtieron en las estrellas (Wright, 2005).)

Los Chamacoco cuentan que al cortarse el puente que unía ambos reinos se ensanchó el universo y ya nunca más cielo y tierra volvieron a juntarse. En las representaciones de Ogwa Flores mostradas en las Figuras anteriores el cielo se halla poblado de seres que se desplazan, interactúan y conviven con animales y plantas de la región superior. Y esta región se ubica tan solo pocos metros por encima de la copa de los árboles más altos de los bosques de la tierra. De las narraciones se desprende, además, que el mismo cielo refleja su color sobre el manto de hojas de la extremidad del follaje de dichos bosques. Al derrumbarse el châro, la zona de unión (de inserción) de esta suerte de pilar cósmico se cerró, transformando al cielo, que era visto como una capa espesa, dura y gris, en una región estratificada y dividida en múltiples niveles. El imaginario Chamacoco, a partir de ese evento fundacional de derrumbe, desarticula la ecúmene en dos mundos que se oponen: el reino de arriba, figurado ahora como un cielo semiesférico; el mundo inferior, apoyado sobre aguas primigenias, imaginado como un mundo náufrago (Sequera, 2006).

Esta separación Cielo-Tierra-Inframundo ya fue documentada ampliamente, tanto en trabajos antropológicos como en otros relacionados con la historia de las religiones. Nuevamente Eliade, en su libro Imágenes y símbolos, menciona el caso de los pigmeos Semang de la península de Malaca (península malaya). En el centro del mundo Semang se ubica una roca enorme (o quizás una colina de piedra caliza; ver Evans, 1968 [1937]), llamada Batu Ribn o Batu 'Rem, que cubre las regiones inferiores y, en tiempos antiguos, constituía la base sobre la que se elevaba un tronco de árbol que llegaba al cielo. Inframundo, centro de la Tierra y la puerta del cielo quedaban entonces unidas por un mismo eje. Este eje, a su vez, era el camino a tomar para transitar entre una y otra región. Este pueblo narraba que en otro tiempo la comunicación con la divinidad y el cielo eran naturales y simples, pero que luego de una falta ritual esta relación se interrumpió. Quedó entonces como privilegio de chamanes lo que antes era natural para todos los integrantes del mundo Semang.

## El espacio-mundo Chamacoco

A partir del derrumbe del châro mitológico, evento de quiebre sucedido en épocas primigenias de la historia de los Ybytóso y los Tomárâho, el universo adquiere una arquitectura particular, donde se separan el mundo de arriba del subterráneo. El primero está ubicado por encima de la superficie terrestre e incluye seis estratos, comenzando con la zona de hábitat usual de las especies vegetales y animales. Esta capa, sólida y seca, es también morada de los hombres, y alcanza una altura equivalente a la de la palma más alta. A esta región se la llama pôrr iut, que en lengua Chamacoco significa cielo más bajo. A continuación se ubica una capa situada por encima de la terrestre y caracterizada por su humedad. Es llamada pôrr pehêt o también pôrr erîch que, en lengua indígena, significa a mitad de los cielos (pehêt significa espacio). Esta capa corresponde a la región donde se ubican las nubes y donde se originan las lluvias. También las tormentas, originadas por los seres osâsero (espíritus de las tormentas), encuentran en esta zona el lugar para desarrollarse [Fig.7]. Esta es la morada de los pajaritos junqueros petîis chyperme (Phleocryptes melanops) que acostumbran volar bajo en zonas pantanosas, y también la de los chamanes guía. Por su parte, los miembros de la comunidad Tomárâho afirman que esta región está poblada de espíritus extraños, llamados osîoro kynaha, seres malignos que transmiten enfermedades, de origen microbiano o infeccioso, endémicas de muchas de esas zonas del Chaco Boreal. El ser mítico Nehmurt también se halla en el pôrr pehét.

El estrato siguiente del mundo de arriba es llamado pôrr pîxt (cielo verdadero) y los Chamacoco lo representan como una capa de neblina espesa, dominio del personaje conocido como Lapyxe, el hacedor de las lluvias. Muchos personajes actúan como guardianes de ciertas cosas, fenómenos y objetos animados o inanimados. Lapyxe, por su parte, es el guardián de las aguas. En esta región de encuentran tanto la Luna como el Sol pero con el segundo ubicado por debajo de la primera. De hecho, el contorno de la Luna marca el límite superior del pôrr pîxt. Este límite constituye el pórtico del cielo, barrera de difícil traspaso, pues es allí donde montan guardia los espíritus extraños osîoro kynaha. Estos son los principales obstáculos que deben sortear los chamanes ahanak cuando intentar viajar por los cielos del mundo de arriba durante sus vuelos chamánicos.

El cuarto cielo se llama pôrr yhyr (cielo alto) y es una ancha zona donde se encuentran las estrellas porrebija, así como también los grupos de estrellas que forman asterismos y constelaciones. Recordemos que en la visión Chamacoco, las estrellas habían surgido a partir de los seres que quedaron rezagados en la región superior al momento de la caída del châro, el árbol cósmico que en época inmemorial unía el cielo y la tierra. La luminaria mayor del cielo nocturno después de la Luna, el planeta Venus, que los indígenas llaman Iohdle, también se halla en este cielo límpido, junto con toda otra especie de objetos celestes. La Vía Láctea, que como vimos es llamada iomyny, se muestra en los cielos diáfanos del Chaco Paraguayo como una notoria franja blanquecina que atraviesa el firmamento, y también se ubica en el pôrr yhyr. Con la excepción de algunos chamanes particularmente visionarios, nadie tiene el poder de incursionar en este cielo lejano y profundo.

Finalmente, los dos últimos estratos del mundo de arriba se llaman pôrr uhur (cielo del horizonte), el primero, y pôrr nahnyk (cielo frío) el más alto. El primero es el umbral del fin del firmamento, y la región en donde surge lo incógnito. El último es visto como el espacio indefinido que va más allá de los cielos interiores, y la región donde se ensancha el universo y predomina lo desconocido. Se trata de un cielo profundo donde el aire no llega. (Cordeu, 1994, ha sugerido una diferente estratificación del mundo superior, más relacionado con las propiedades cromáticas y atmosféricas del cielo.)

Viajemos ahora en la dirección vertical opuesta, esto es, hacia el reino de las profundidades. El mundo subterráneo, surgido en el imaginario Chamacoco luego de la caída del chôro mitológico, se divide en tres estratos principales. Se trata de un mundo oculto y profundo, que se prolonga hacia las entrañas de la tierra y que tiene una constitución viscosa. Se piensa que es una zona donde reina la destrucción. La primera región, nîogoro urr, es una zona de humedales y cursos de agua superficiales. En este medio fluído se desplaza la mítica anguila dyhylygyta y también el uriche, similar al actual lobito de río (de la especie Lontra longicaudis). Estos animales coexisten también con espíritus que toman la apariencia de peces. Serían estos últimos quienes otorgarían poderes para luchar contra los espíritus extraños osîoro kynaha a los chamanes.

A la segunda capa subterránea se la imagina como una zona de aguas profundas, mezclada con lodo espeso y viscoso. Se la conoce como hñymich yhyrt, literalmente tierra de monte alto, y es la morada de monstruos similares a anguilas, entre los cuales el más fabuloso tanto por su dimensión como por su cabeza gigante y rojiza, es el que los nativos llaman pêeta. Esta anguila-monstruo concede a ciertos chamanes el poder para desplazarse ágilmente por esta capa subterránea, para luego emerger del mundo náufrago a gran velocidad por cualquier parte de la tierra. Por último, la tercera capa subterránea se llama hñymich urruo y significa bajo tierra. En esta región se localiza el mundo oscuro y putrefacto de los cadáveres y la morada del ser llamado amyrmy lata, de aspecto similar a un armadillo gigante (Priodontes maximus) y al que se vincula con los chamanes iniciados.

Si bien estos diferentes estratos del mundo náufrago invocan destrucción, muerte y finalmente putrefacción de todo lo vivo, pueden sin embargo liberar ciertas fuerzas de ascensión como las que impulsan a los chamanes en la segunda capa (Sequera, 2006) y también las que caracterizan a los ahnapsûro (o axnábsero), los seres míticos originarios del mundo Chamacoco [Fig.8]. Descritos por primera vez por Boggiani en el año 1900 (Baldus, 1932; Susnik 1995 [1969]), los ahnapsûro son mencionados por todos los Chamacoco, aunque sólo los Tomárâho mantienen aún hoy en día la práctica ritual de la representación mítica (los mitos de origen emuhno). Estos seres acuáticos, de inmenso poderío y cuyo cuerpo está completamente recubierto de escamas y plumas, son los fundadores de la cultura Chamacoco, y en los orígenes convivieron con los primeros habitantes yxyro en armonía y les enseñaron a buscar el alimento para alimentar a la tribu. Su aparición hoy en el campamento Tomárâho es fuente de estremecimiento y alteración de la calma de la comunidad; regularmente se organizan representaciones rituales que conjuran ese temor [Fig.9].

## Perspectiva y proyectos futuros

Se mostraron algunas pinceladas de la inmensa riqueza de los Chamacoco en lo que hace a su relación con la naturaleza y con el cielo. Un largo y paciente trabajo etnográfico ha logrado revelar las características más salientes de este grupo aborigen en su conjunto y de las parcialidades Tomárâho e Ybytóso en particular. La familiarización con su cultura, la transcripción de su lengua, el lento mejoramiento de su situación social, y la educación respetando sus tradiciones e identidad cultural, son todos elementos en los que se ha ido progresando significativamente.

En este importante trabajo etnográfico, sin embargo, los aspectos astronómicos no fueron abordados con igual intensidad que los demás. Por ello, resta la tarea de hacer un trabajo detallado de reconocimiento e identificación de aspectos salientes del cielo Chamacoco, como así también de la verdadera significación del chäro, el árbol cósmico que unía el cielo con la tierra y que aún hoy es central en los rituales del origen del mundo [Fig.10].

Estrellas prominentes visibles en diferentes momentos del año (como Sirio, o incluso asterismos notorios como el hoy conocido de las Tres Marías en el cinturón de Orión), sus nombres y las historias que seguramente se narraban sobre ellas; su imaginario sobre la presencia y características de la Vía Láctea; sus interpretaciones sobre fenómenos asombrosos y sorpresivos como los eclipses totales de Sol, sobre apariciones prolongadas en el cielo como las de los cometas o incluso esporádicas, como las estrellas fugaces, son algunos aspectos que, creemos, requieren una mayor atención.

La historia de Iohdle (Venus) que muchos años atrás vivió una relación matrimonial con un joven gentil Tomárâho, es tan solo una de las muchas narraciones con elementos astronómicos que acostumbraban contar los ancianos. Muchas de estas historias aún perduran en la mente de los representantes de mayor edad de este grupo y un proyecto necesario es poder incorporarlas al patrimonio inmaterial de la humanidad antes de que se extravíen en las insondables nubes del tiempo. Otro estudio que vemos en el horizonte es tratar de interpretar la relación que existe entre las técnicas vocales e instrumentales de los Chamacoco y su visión del mundo, en especial en el empleo que ellos hacen de las sonajas (paîkâra) que, como vimos, vienen a representar el cielo estrellado.

Asimismo, nuestro proyecto pretende llevar a cabo, junto a integrantes de la comunidad Tomárâho actual, un trabajo de colecta de datos y reconocimiento de distribuciones de estrellas y constelaciones, zonas oscuras del cielo, nebulosas, etc. Serían mapas interpretados, con una carga simbólica muy importante, del cielo Tomárâho para la región del alto Chaco en diferentes épocas del año, que quizás nos digan mucho sobre el imaginario aborigen y sobre los elementos de la vida cotidiana que ellos proyectaban en el fondo oscuro del cielo. En resumen, pensamos que la representación Tomárâho del cielo no fue aun suficientemente explorada como un objeto de estudio en sí mismo y que se requiere más trabajo de campo en el área de la etnoastronomía como una actividad interdisciplinaria que incluya las competencias de antropólogos y de astrónomos.

# Bibliografía

## Figuras referenciadas en el texto (leyendas de las figuras):

**Fig.1**: Indígenas del Paraguay actual. Fuente: M. Chase-Sardi (infografía proveniente del Museo del Barro, Asunción del Paraguay).

**Fig.2**: Mapa de la región del Alto Chaco paraguayo, con algunas ciudades y pueblos destacados. Fuente: The curse of Nemur, de Ticio Escobar.

**Fig.3**: Sonajas empleadas por los chamanes Chamacoco, cuya parte superior es identificada con el centro de la bóveda celeste.

**Fig.4**: El ñandú, llamado Pemme-Kamytêrehe por los Chamacoco, se encarga del cuidado de Deich, el Sol.

**Fig.5**: El chãro o kululte, árbol-sostén del mundo en la visión Chamacoco, nexo entre la tierra y el cielo.

**Fig.6**: El chãro como pilar y sostén del mundo, en épocas previas a su caída por la acción de Dagylta.

**Fig.7**: Los espíritus de las tormentas, osâsero, en los dibujos de Ogwa Flores Balbuena.

**Fig.8**: Los seres míticos originarios del mundo Chamacoco, llamados ahnapsûro.

**Fig.9**: Representaciones rituales representando a los seres míticos originarios ahnapsûro.

**Fig.10**: Lugar prominente del chãro durante la representación del origen del mundo Tomárâho.

**Figura 1.**

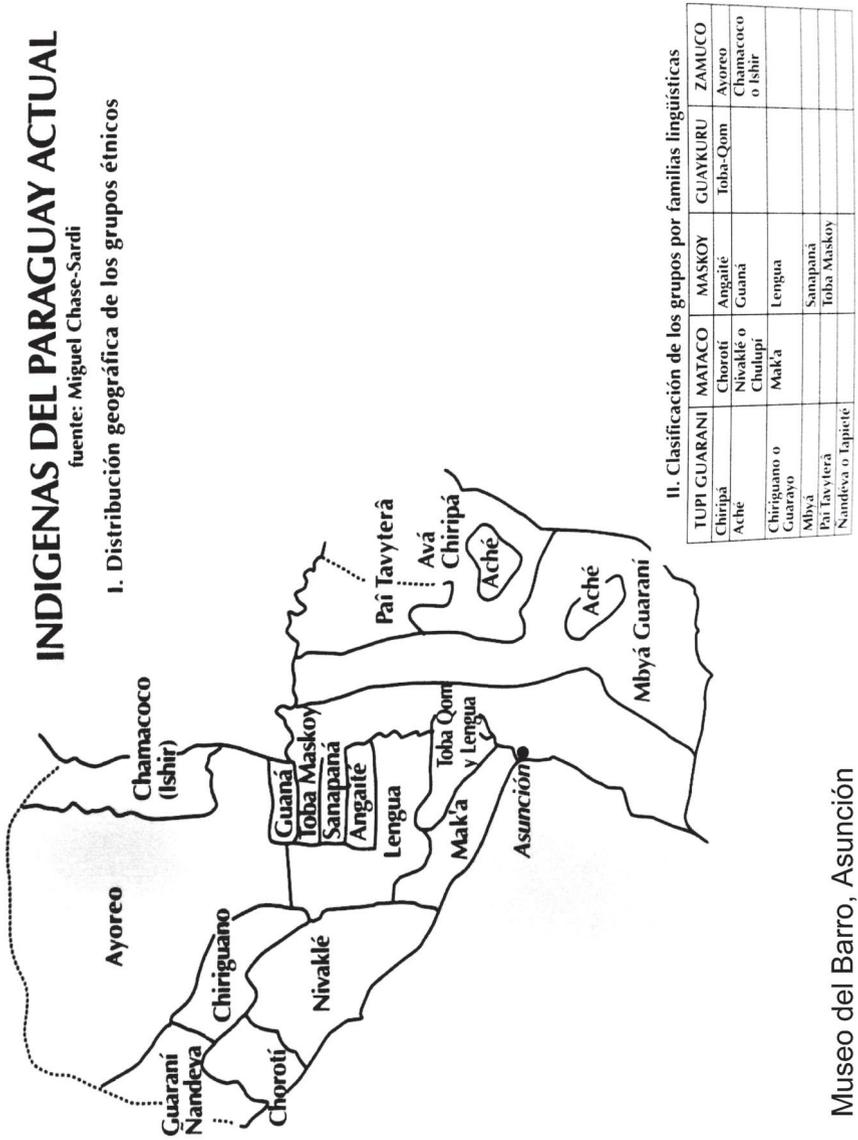

**Figura 2.**

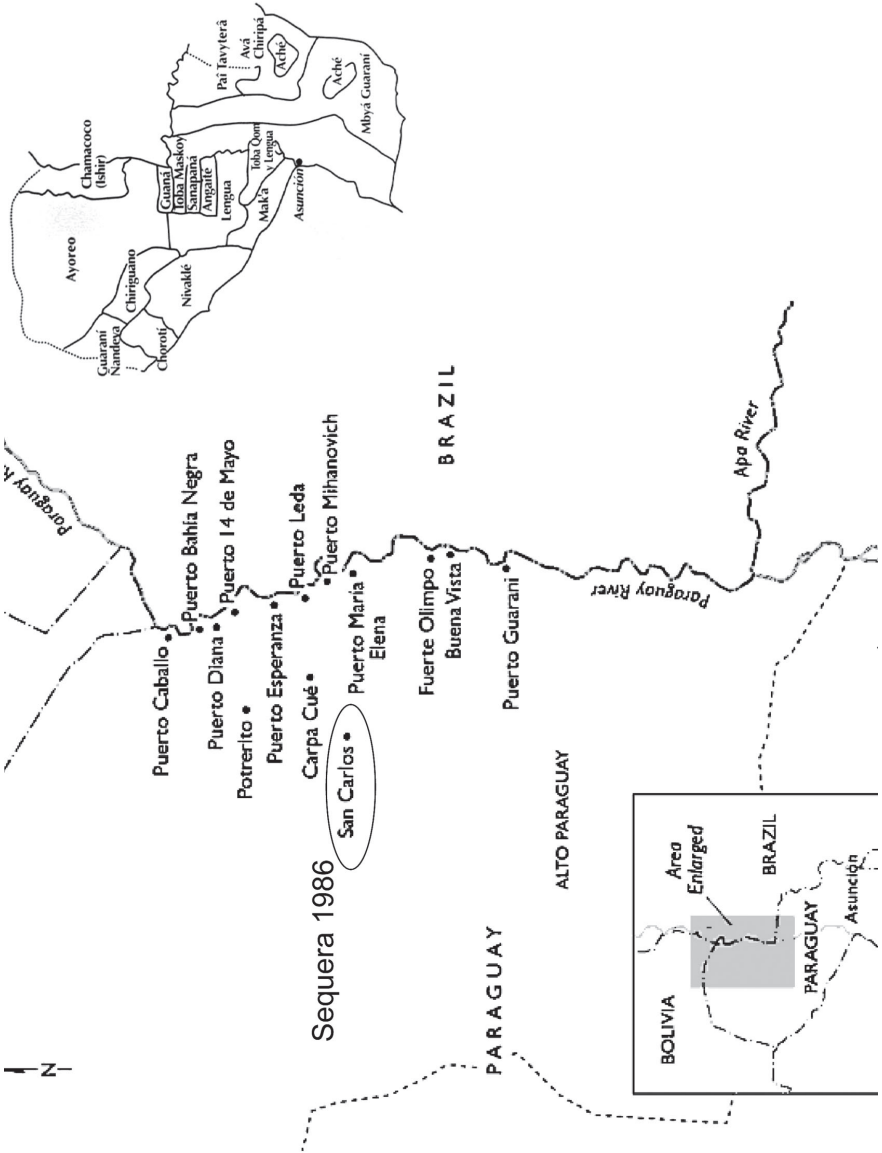

**Figura 3.**

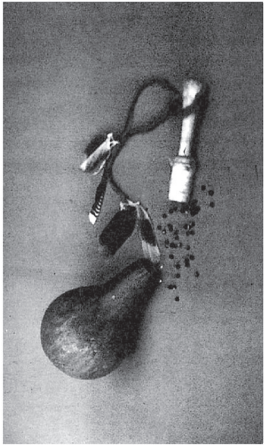
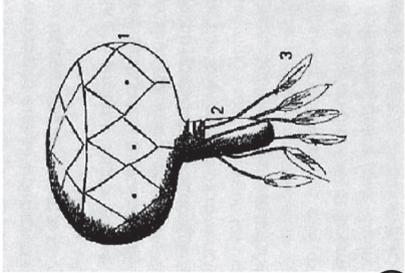
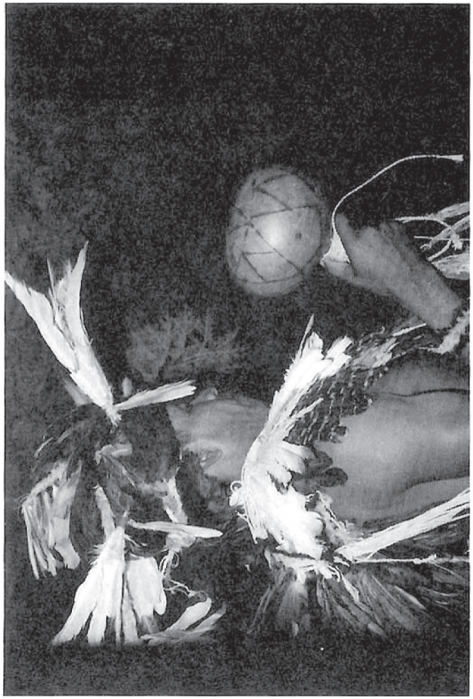

*Mahâra Dohorâta Wylky*, chamán estelar interpretando sus cantos de repertorio individual, y engalanado con la exuberancia de su arte plumario. Foto: G. Sequera. Represa primera, Alto Paraguay, 1986.

chamanismo chamacoco

sonaja: **osecha** o **paîkâra** (calabaza *Lagenaria siceraria*)

**Figura 4.**

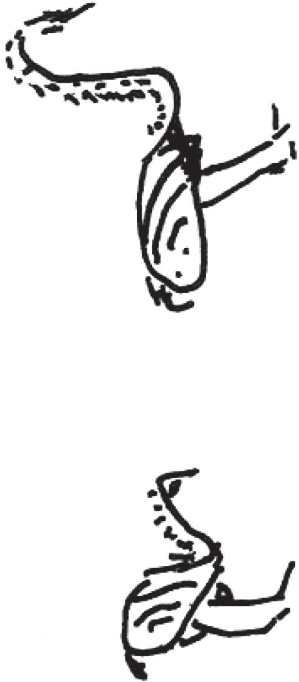

Pemme-Kamytêrehe

Rhea americana (ñandú), es el "ser protector" de Deich, el Sol.
Dibujos de Wulky Dohorâta (chamán estelar Tomárâho, Peichota, 1991)

**Figura 5.**

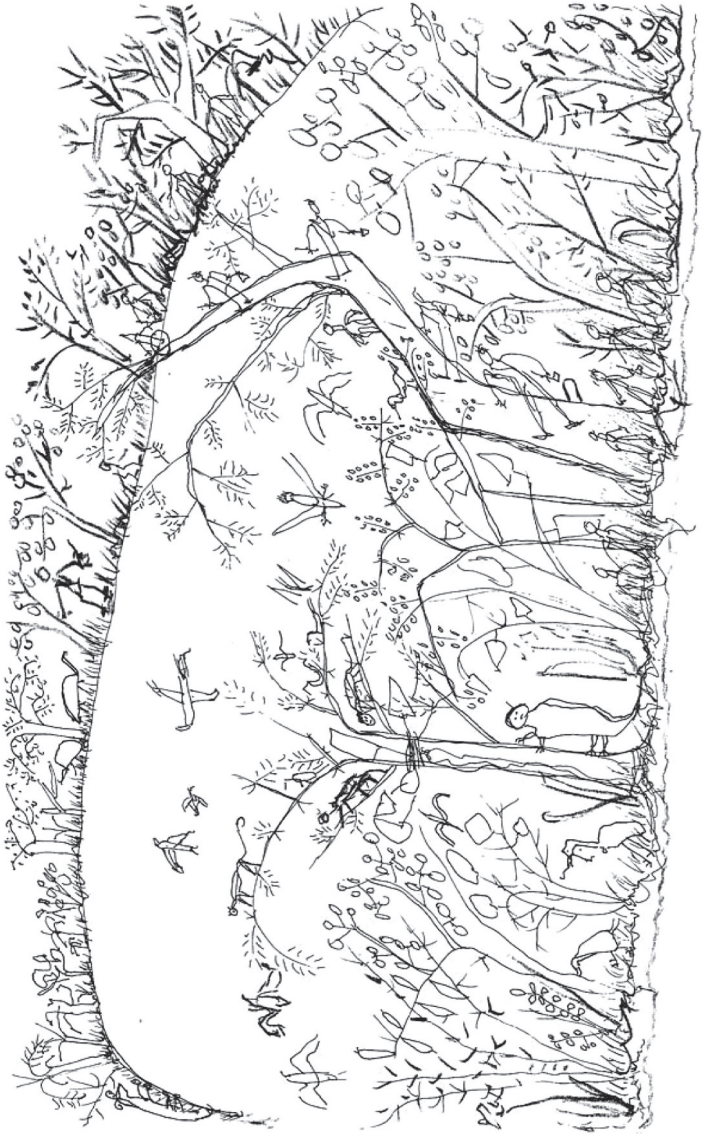

Dibujo de Ogwa Flores Balbuena, integrante de la comunidad Ybytóso, 1991

**Figura 6.**

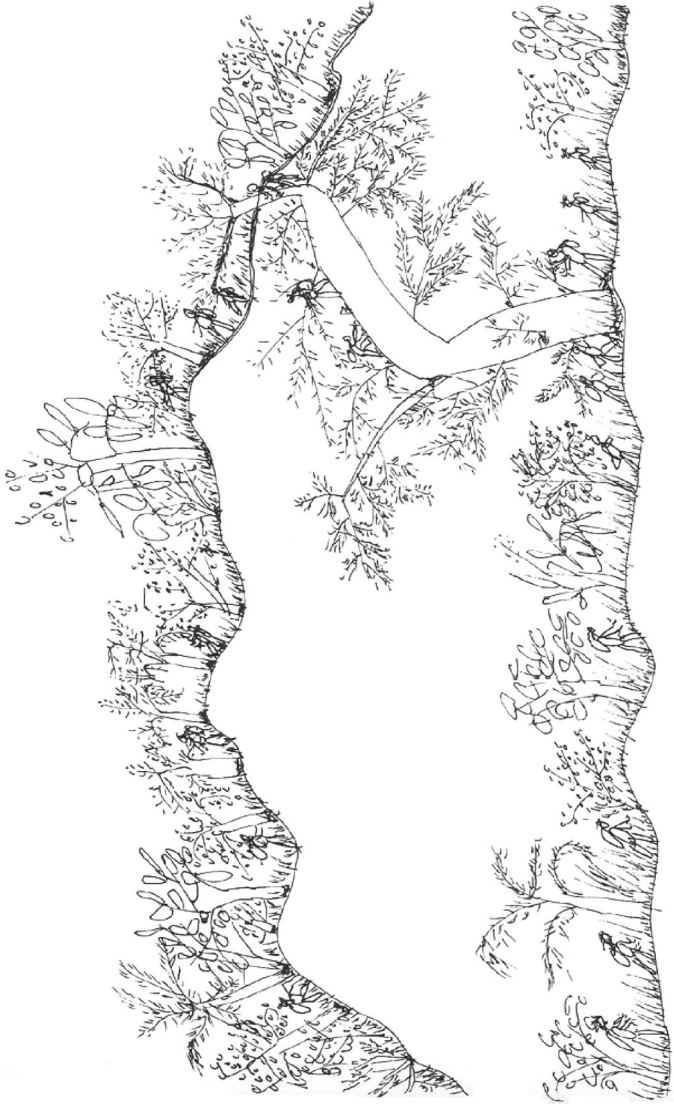

Dibujo de Ogwa Flores Balbuena, integrante de la comunidad Ybytóso, 1988

**Figura 7.**

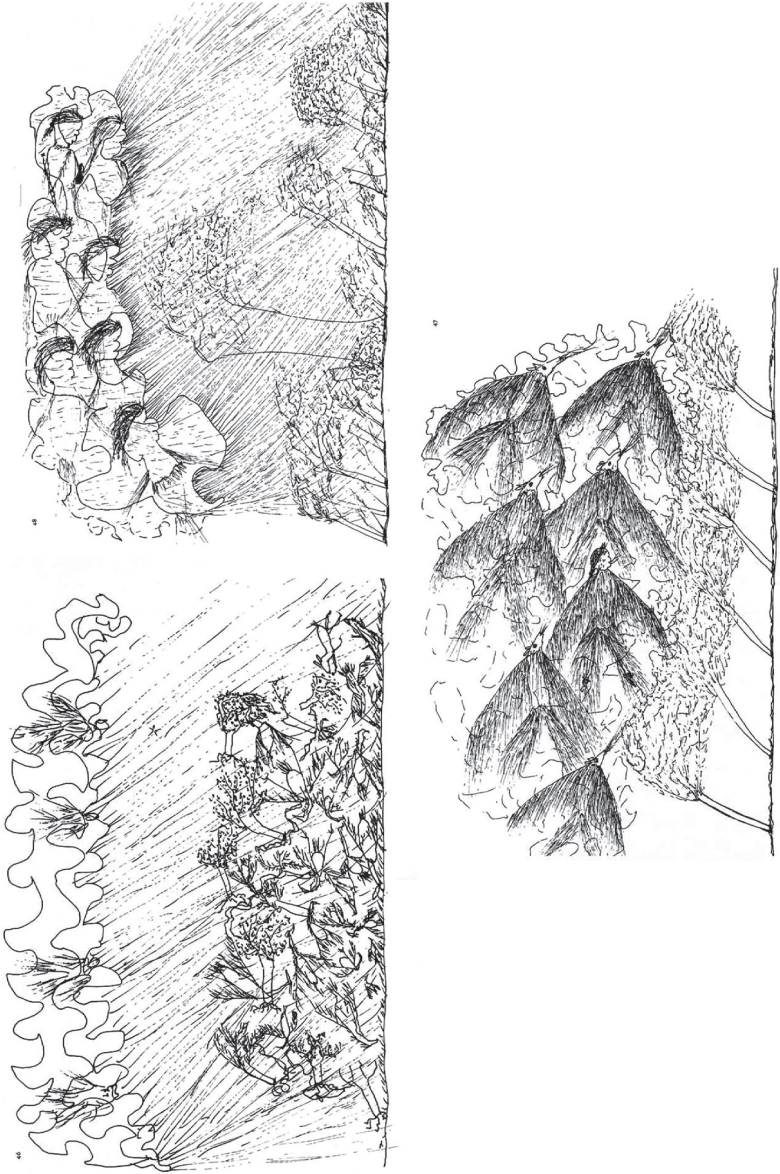

Osâsero, espíritus de las tormentas
Dibujo de Ogwa Flores Balbuena (Ybytóso), 1990, 1992, 1993

**Figura 8.**

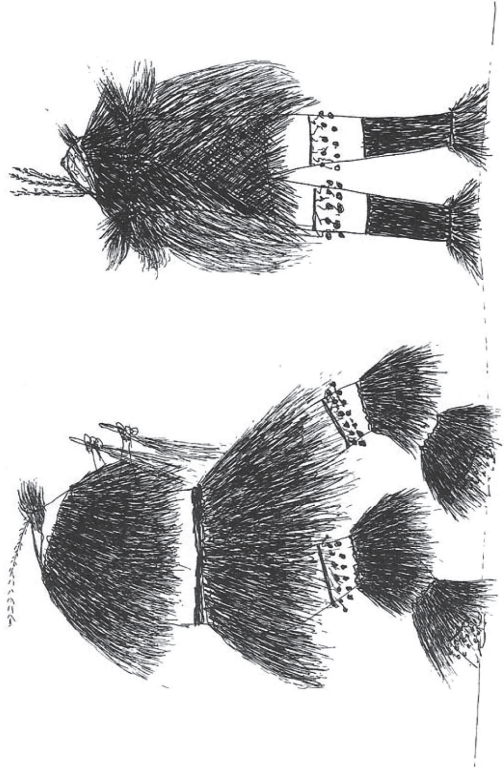

Seres míticos originarios *ahnapsŭro*.
Dibujo de Ogwa Flores Balbuena (Ybytóso), 1992

**Figura 9.**

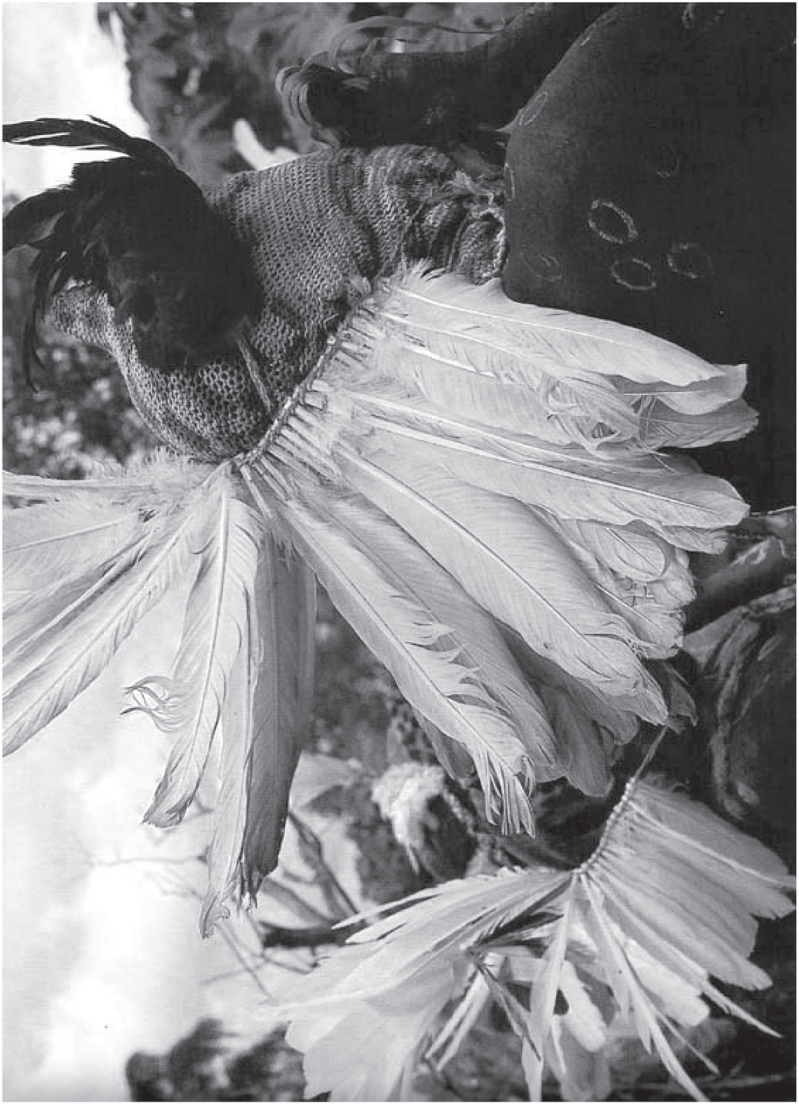

Seres míticos originarios *ahnapsûro* en el escenario principal del espacio comunitario Tomárâho, G.Sequera, Peichota, 1984

## Figura 10.

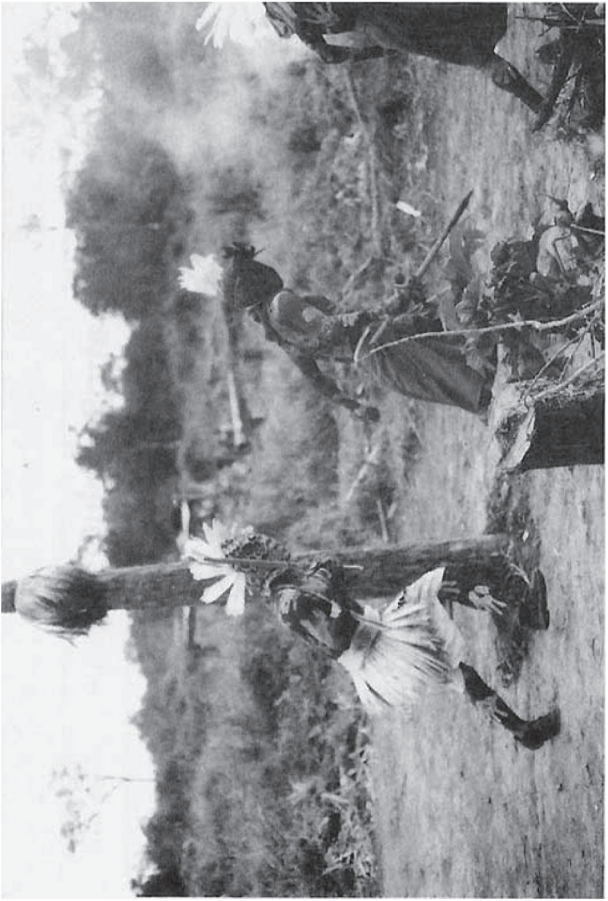

Ritual del Origen del Mundo.
Espacio comunitario Tomáraho, G.Sequera, Peichota, 1988